\documentstyle[graphics,fullpage,epsf,twocolumn]{article}

\twocolumn

\title{{\em Ab Initio} Study of Screw Dislocations in Mo and Ta: A new
picture of plasticity in bcc transition metals}
\author{Sohrab Ismail-Beigi$^\dag$ and T.A.~Arias$^\ddag$\\
$\dag$ Department of Physics, Massachusetts Institute of Technology,
Cambridge, MA 02139\\
$\ddag$ Laboratory of Atomic and Solid State Physics, Cornell University,
Ithaca, NY 14853}

\date{\ }

\begin{document}

\maketitle

\begin{abstract}
We report the first {\em ab initio} density-functional study of
$\langle 111\rangle$ screw dislocations cores in the bcc transition
metals Mo and Ta. Our results suggest a new picture of bcc plasticity
with symmetric and compact dislocation cores, contrary to the
presently accepted picture based on continuum and interatomic
potentials.  Core energy scales in this new picture are in much better
agreement with the Peierls energy barriers to dislocation motion
suggested by experiments.
\end{abstract}

\vspace{.5cm}

The microscopic origins of plasticity are far more complex and less
well understood in bcc metals than in their fcc and hcp counterparts.
For example, slip planes in fcc and hcp metals are almost invariably
close-packed, whereas in bcc materials many slip systems can be
active.  Moreover, bcc metals violate the Schmid law that the
resistance to plastic flow is constant and independent of slip system
and applied stress\cite{Schmid}.

Detailed, microscopic observations have established that in bcc metals
at low temperatures, long, low-mobility $\langle 111\rangle$ screw
dislocations control the plasticity\cite{Vitek,Duesbery}.  Over the
last four decades, the dominant microscopic picture of bcc plasticity
involves a complex core structure for these dislocations.  The key
ingredient of this intricate picture is an extended, non-planar
sessile core which must contract before it moves.  The first such
proposed structure respected the symmetry of the underlying lattice
and extended over many lattice constants\cite{Hirsch}.  More recent
and currently accepted theories, based on interatomic potentials,
predict extension over several lattice constants and spontaneously
broken lattice symmetry\cite{Vitek,Duesbery,MGPT,MGPTTa}.  While these
models can explain the overall non-Schmid behavior, their predicted
magnitude for the critical stress required to move dislocations ({\em
Peierls stress}) is uniformly too large by a factor of about three
when compared to experimental yield stresses extrapolated to
zero-temperature\cite{Duesbery,Duesbery2}.

We take the first {\em ab initio} look at dislocation core structure
in bcc transition metals.  Although we study two metals with quite
different mechanical behavior, molybdenum and tantalum, a consistent
pattern emerges from our results which, should it withstand the test
of time, will require rethinking the presently accepted picture.
Specifically, we find screw dislocation cores with compact structures,
without broken symmetry, and with energy scales which appear to be in
much better accord with experimental Peierls barriers.

{\em Ab initio methodology -- } Our {\em ab initio} calculations for
Mo and Ta are carried out within the total-energy plane-wave density
functional pseudopotential approach\cite{RMP}, using the
Perdew-Zunger\cite{PerdewZunger} parameterization of the
Ceperly-Alder\cite{CeperlyAlder} exchange-correlation energy.
Non-local pseudopotentials of the Kleinman-Bylander form\cite{KB} are
used with $s$, $p$, and $d$ channels.  The Mo potential is optimized
according to\cite{Rappe} and the Ta potential is from\cite{Tapot}.  We
use plane wave basis sets with energy cutoffs of 45 Ryd for Mo and 40
Ryd for Ta to expand the wave functions of the valence (outermost $s$
and $d$) electrons.  Calculations in bulk show these cutoffs to give
total system energies to within 0.01 eV/atom.  We carry out electronic
minimizations using the analytically continued approach\cite{ACPRL}
within the DFT++ formalism\cite{DFT++}.

To gauge the reliability of the pseudopotentials,
Table~\ref{table:bulkprops} displays our {\em ab initio} results for
the materials' lattice constants and those elastic moduli most
relevant for the study of $\langle 111\rangle$ screw dislocations.
The tabulated moduli describe the long-range elastic fields of the
dislocations ($K$), the coupling of displacement gradients along the
dislocation axis $z$ to core-size changes in the orthogonal $x,y$
plane ($c_{xx,zz}=(c_{11}+5c_{12}-2c_{44})/6$), and the coupling of
core-size changes to themselves in the plane
($c_{xx,xx}=(c_{11}+c_{12}+2c_{44})/2$ and
$c_{xx,yy}=(c_{11}+2c_{12}-2c_{44})/3$).  These results indicate that
our predicted core energy differences should be reliable to within
better than \mbox{$\sim30$\%}, which suffices for the purposes of our
study.

{\em Preparation of dislocation cells --} The cell we use for
dislocation studies has lattice vectors $\vec{a}_1 = 5a[1,-1,0]$,
$\vec{a}_2=3a[1,1,-2]$, and $\vec{a}_3=a[1,1,1]/2$, where $a$ is the
lattice constant.  We call this ninety-atom cell the ``5$\times$3''
cell in reference to the lengths of $\vec{a}_1$ and $\vec{a}_2$, and the
Burgers vectors of all of the dislocations in our work are along
$\vec{a}_3$.  Eight $k$-points $k_1=k_2={1 \over 4},k_3\in\pm\{{1\over
16},{3\over 16},{5\over 16},{7\over 16}\}$ sample the Brillouin zone
in conjunction with a non-zero electronic temperature of
$k_BT=0.1$~eV, which facilitates the sampling of the Fermi surface.
These choices give total energies to within 0.01 eV/atom.

Given the relatively small cell size, we wish to minimize the overall
strain and the effects of periodic images.  We therefore follow
\cite{PayneSi} and employ a quadrupolar arrangement of dislocations (a
rectangular checkerboard pattern in the $\vec{a}_1,\vec{a}_2$ plane).
This ensures that dislocation interactions enter only at the
quadrupolar level and that the net force on each core is zero by
symmetry, thereby minimizing perturbations of core structure due to the
images.  As was found in\cite{PayneSi} and as we explore in detail
below, we find very limited impact of finite-size effects on the
cores when following this approach.

In bcc structures, screw dislocations are known to have two
inequivalent core configurations, termed ``easy'' and
``hard''\cite{Vitek,MGPT,MGPTTa}.  These cores can be obtained from
one another by reversing the Burgers vector of a dislocation line
while holding the line at a fixed position.  We produce cells with
either only easy or only hard cores in this way.  To create atomic
structures for the cores, we proceed in three stages.  First, we begin
with atomic positions determined from isotropic elasticity theory for
our periodic array of dislocations.  Next, we relax this structure to
the closest local energy minimum within the interatomic MGPT model for
Mo\cite{MGPT}.  Since we do not have an interatomic potential for Ta
and expect similar structures in Ta and Mo\cite{MGPTTa}, we create
suitable Ta cells by scaling the optimized MGPT Mo structures by the
ratio of the materials' lattice constants.  Finally, we perform
standard {\em ab initio} atomic relaxations on the resulting MGPT
structures until all ionic forces in all axial directions are less
than 0.06 eV/\AA.

{\em Extraction of core energies -- } The energy of a long, straight
dislocation line with Burgers vector $\vec b$ is $E = E_c(r_c) +
Kb^3\ln(L/r_c)$ per $b$ along the line\cite{HirthLothe}, where $L$ is
a large-length cutoff, and $K$ is an elastic modulus (see Table
\ref{table:bulkprops}) computable within anisotropic elasticity
theory\cite{Head}. The core radius $r_c$ is a short-length cutoff
inside of which the continuum description fails and the discrete
lattice and electronic structure of the core become important.
$E_c(r_c)$ measures the associated ``core energy'', which, due to
severe distortions in the core, is most reliably calculated by {\em ab
initio} methods.

The energy of our periodic cell contains both the energy of four
dislocation cores and the energy stored outside the core radii in the
long-range elastic fields.  To separate these contributions, we start
with the fact that two straight dislocations at a distance $d$ with
equal and opposite Burgers vectors have an anisotropic elastic energy
per $b$ given by $E = 2E_c(r_c) + 2Kb^3\ln(d/r_c)$.  Next, by
regularizing the infinite sum of this logarithmically divergent pair
interaction, we find that the energy per dislocation per $b$ in our
cell is given by
\begin{equation}
E = E_c(r_c) + Kb^3\left[\ln\left({|\vec{a}_1|/2\over r_c}\right) +
A\left({|\vec{a}_1|\over |\vec{a}_2|}\right)\right].
\label{eq:ewaldsum}
\end{equation}
The function $A(x)$ contains all the effects of the infinite
Ewald-like sums of dislocation interactions and has the value $A =
-0.598\,846\,386$ for our cell. Subtracting the long-range elastic
contribution (the second term of (\ref{eq:ewaldsum})) from the total
energy, we arrive at the core energy $E_c$.

To test the feasibility of this approach, we compare $E_c(r_c)$ for
the MGPT potential as extracted with the above procedure from cells of
two different sizes: the 5$\times$3 cell and the corresponding
9$\times$5 cell.  (The MGPT is {\em fit} to reproduce experimental
elastic moduli, so $K$ is given in Table \ref{table:bulkprops}.)  With
the choice $r_c=2b$, Table~\ref{table:MGPT5x39x5} shows that our
results, even for the 5$\times$3 cell, compare quite favorably with
those of\cite{MGPT,MGPT2}, especially given that our 5$\times$3 and
9$\times$5 cells contain only ninety and 270 atoms respectively,
whereas the cited works used cylindrical cells with a single
dislocation and {\em two thousand} atoms or more.  Given the
suitability of the 5$\times$3 cell, all {\em ab initio} results
reported below are carried out in this cell.

{\em Ab initio core energies -- } Except for the Mo hard core, all the
core structures relax quite readily from their MGPT configurations to
their equilibrium {\em ab initio} structures.  The Mo hard-core
configuration, however, spontaneously relaxes into easy cores,
strongly indicating that the hard core, while meta-stable within MGPT
by only 0.02~eV/$b$\cite{MGPT2}, is not stable in density functional
theory.  We do not believe that this instability is due to finite-size
effects, which appear to be quite small for the reasons outlined
previously.

Table~\ref{table:compareabiMGPT} compares our {\em ab initio} results
to available MGPT results for core energies in Mo and Ta.  To make
comparison with the MGPT, for the unstable Mo hard core we evaluate
the {\em ab initio} core energy at the optimal MGPT atomic
configuration (column AI$^*$ in Table~\ref{table:compareabiMGPT}).
Note that, in computing hard--easy core energy differences, the
long-range elastic contributions cancel so that these differences are
much better converged than the absolute core energies.

\begin{figure}
\epsfxsize=3.0in
\epsfbox{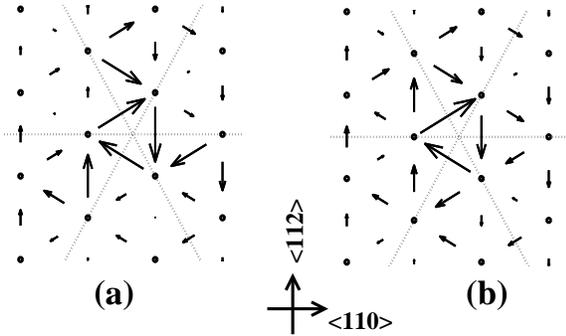}
\caption{DD maps as found in the MGPT model for (a) easy, and (b) hard
Mo dislocation cores.  Dotted lines indicate axes of $C_2$ symmetry of
the $D_3$ symmetry group.}
\label{fig:MGPTDDmaps}
\end{figure}

Table~\ref{table:compareabiMGPT} shows that the MGPT hard--easy core
energy differences are much larger than the corresponding {\em ab
initio} values by approximately a factor of three.  The accuracy of the
elastic moduli of Table \ref{table:bulkprops}, combined with the high
transferability of the local-density pseudopotential approach,
indicates that this factor of three is not an artifact of our
approximations.  We believe that the reason for this discrepancy is
that the MGPT is less transferable. Having been forced to fit {\em
bulk} elastic moduli and thus long-range distortions, the MGPT may not
describe the short wavelength distortions in the cores with high
accuracy.  An examination of Mo phonons along $[100]$ provides
poignant evidence: the MGPT frequencies away from the zone center are
too large when compared to experimental and band-theoretic
values\cite{MGPT} and translate into spring constants that are up to
approximately three times too large.

The magnitude of the core energy difference has important implications
for the magnitude of the Peierls energy barrier and Peierls stress for
the motion of screw dislocations in Mo and Ta.  In a recent Mo MGPT
study\cite{MGPT2}, the most likely path for dislocation motion was
identified to be the $\langle 112\rangle$ direction: the moving
dislocation core changes from easy to hard and back to easy as it
shifts along $\langle 112\rangle$.  The energy barrier was found to be
$0.26$ eV/$b$, very close to the MGPT hard--easy energy difference
itself.  The fact that the {\em ab initio} hard--easy energy
differences in Mo and Ta are smaller by about a factor of three than
the respective interatomic values suggests that the {\em ab initio}
energy landscape for the process has a correspondingly smaller scale.
If so, the Peierls stress in Mo and Ta should also be correspondingly
smaller and in much better agreement with the values suggested by
experiments.

{\em Dislocation core structures -- } Figure~\ref{fig:MGPTDDmaps}
shows differential displacement (DD) maps\cite{Vitek} of the core
structures we find in our ninety-atom supercell when working with the
interatomic MGPT potential for Mo.  Our DD maps show the atomic
structure projected onto the (111) plane.  The vector between a pair
of atomic columns is proportional to the change in the $[111]$
separation of the columns due to the presence of the dislocations.
The maps show that both easy and hard cores have approximate 3-fold
rotational ($C_3$) point-group symmetry about the out-of-page $[111]$
axis through the center of each map.  The small deviations from this
symmetry reflect the weakness of finite-size effects in our
quadrupolar cell.  The hard core has three additional 2-fold
rotational ($C_2$) symmetries about the three $\langle 110\rangle$
axes marked in the maps, increasing its point-group symmetry to
the dihedral group $D_3$ which is shared by the underlying crystal.
The easy core, however, shows a strong spontaneous breaking of this
symmetry: its core spreads along only three out of the six possible
$\langle 112\rangle$ directions.  Our results reproduce those
of\cite{MGPT,MGPT2} who employed much larger cylindrical cells with
open boundaries, underscoring the suitability of our cell for
determining core structure.  This symmetry-breaking core extension is
that which has been theorized to explain the relative immobility of
screw dislocations and violation of the Schmid law in bcc metals.

\begin{figure}
\epsfxsize=3.0in
\epsfbox{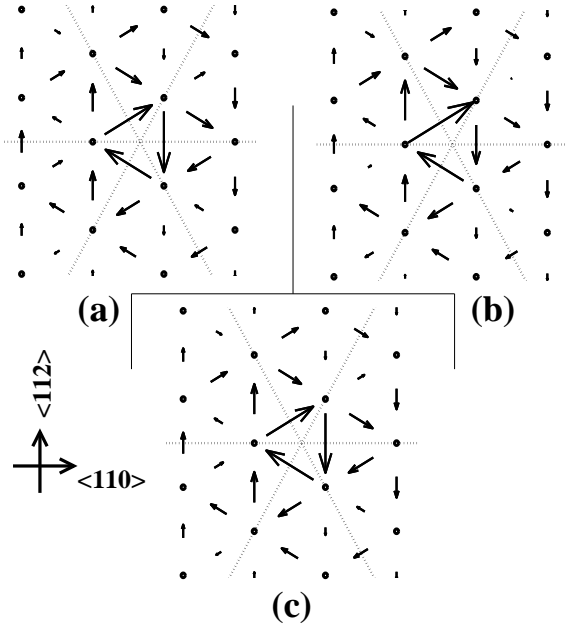}
\caption{DD maps of the {\em ab initio} dislocation cores: (a) Ta
easy, (b) Ta hard; and (c) Mo easy. Dotted lines indicate axes of
$C_2$ symmetry of the $D_3$ symmetry group.}
\label{fig:abiDDmaps}
\end{figure}

Figure~\ref{fig:abiDDmaps} displays DD maps of our {\em ab initio}
core structures.  Contrary to the atomistic results, we find that the
low-energy easy cores in Mo and Ta have full $D_3$ symmetry and do not
spread along the $\langle 112\rangle$ directions.  Combining this with
the above results concerning core energetics, we have two examples for
which our pseudopotentials are sufficiently accurate to disprove the
conventional wisdom that generic bcc metallic systems {\em require}
broken symmetry in the core to explain the observed immobility of
screw dislocations.

Turning to the hard core structures, the {\em ab initio} resuts for Ta
show a significant distortion when compared to the atomistic core
(contrast Figure~\ref{fig:MGPTDDmaps}b and
Figure~\ref{fig:abiDDmaps}b). As the {\em ab initio} Mo hard core was
unstable, we believe that this distortion of the Ta hard core suggests
that this core is much less stable within density functional theory
than in the atomistic potentials.

\begin{figure}
\epsfxsize=3.0in
\epsfbox{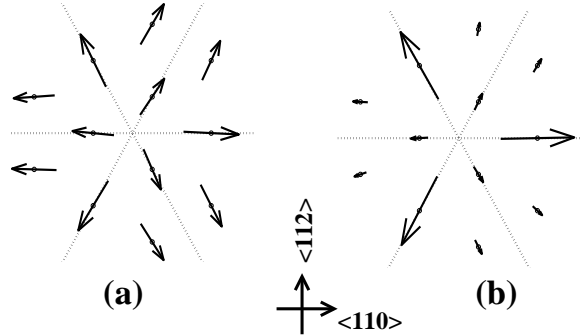}
\caption{Planar displacement maps of the {\em ab initio} (a) Mo easy,
and (b) Ta easy cores. Vectors show in-plane (111) atomic shifts and
have been magnified by a factor of fifty. Dotted lines indicate the
$C_2$ axes of the $D_3$ symmetry group.}
\label{fig:horizmaps}
\end{figure}

To complete the specification of the three-dimensional {\em ab initio}
structure of easy cores in Mo and Ta, Figure~\ref{fig:horizmaps}
presents maps of the atomic displacement in the (111) plane.  The
small atomic shifts, which are due entirely to anisotropic effects,
are shown as in-plane vectors centered on the bulk atomic positions and
magnified by a factor of {\em fifty}.  To reduce noise in the figure,
before plotting we perform $C_3$ symmetrization of the atomic
positions about the $[111]$ axis passing through the center of the
figure.  As all the dislocation cores in our study have a minimum of
$C_3$ symmetry, this procedure does not hinder the identification of
possible spontaneous breaking of the larger $D_3$ symmetry group.  Our
maps indicate that the easy cores in both Mo and Ta have full $D_3$
symmetry.

Recent high-resolution electron microscopy explorations of the
symmetry of dislocations in Mo have focused on the small shifts in the
(111) plane of columns of atoms along $[111]$\cite{Sigle}.  This
pioneering work reports in-plane displacements extending over a range
much greater than the corresponding MGPT results and also much greater
than what we find {\em ab initio}.  In\cite{Sigle} this is attributed
to possible stresses from thickness variations and foil bending.  We
believe this makes study of the internal structure of the core
difficult, and that cleaner experimental results are required to
resolve the nature of the symmetry of the core and its extension.

In conclusion, our first principles results show no preferential
spreading or symmetry breaking of the dislocation cores and exhibit an
energy landscape with the proper scales to explain the observed
immobility of dislocations.  Atomistic models which demonstrate core
spreading and symmetry breaking, both of which tend to reduce the
mobility of the dislocations, are well-known to over-predict the
Peierls stress.  The combination of these two sets of observations
argues strongly in favor of much more compact and symmetric bcc screw
dislocation cores than presently believed.

This work was supported by an ASCI ASAP Level 2 grant (contract
\#B338297 and \#B347887).  Calculations were run primarily at the
Pittsburgh Supercomputing Center with support of the ASCI program and
also on the MIT Xolas prototype SMP cluster.  We thank members of the
H-division at Lawrence Livermore National Laboratories for providing
the Ta pseudopotential, the Mo MGPT code, and many useful discussions.

\begin{table*}
\begin{tabular}{c||c|c|c||c|c|c}
& \multicolumn{3}{c||}{Mo} & \multicolumn{3}{|c}{Ta}\\
\hline
&\ AI \ & \ Expt\ & \ Error \ & \ AI \ & \ Expt\ & \ Error \ \\
\hline
\hline
\ $a$ \  & 3.10 & 3.15 & -1.6\% & 3.25 & 3.30 & -1.5\%\\
\hline
\ $K$\ \ & 1.60 & 1.36 & 18\% & 0.65 & 0.62 & 5\%\\
\hline
\ $c_{xx,zz}$ \ & 2.17 & 1.91 & 14\% & 1.72 & 1.39 & 24\% \\
\hline
\ $c_{xx,xx}$ \ & 5.48 & 4.25 & 29\% & 3.02 & 2.98 & 1.3\%\\
\hline
\ $c_{xx,yy}$ \ & 2.21 & 1.77 & 25\% & 1.72 & 1.49 & 15\%\\
\end{tabular}
\caption{
Lattice constants $a$ (\AA) and elastic moduli (Mbar) for Mo
and Ta based on {\em ab initio} pseudopotentials (AI) and experiments
(Expt)\protect\cite{Tapot}.}
\label{table:bulkprops}
\end{table*}

\begin{table*}
\begin{tabular}{c||c|c|c}
\ \ $E_c$ (eV/$b$) \ \ &\ \ \ \ 5$\times$3\ \ \ \ & \ \ \ \ 9$\times$5\ \ \ \ &
Cylindrical\cite{MGPT,MGPT2}\\
\hline
\hline
hard & 2.57 & 2.57 & 2.66 \\
\hline
easy & 2.35 & 2.31 & 2.42 \\
\hline
\hline
$\Delta$ & 0.22 & 0.26 & 0.24
\end{tabular}
\caption{Core energies for $r_c=2b$ as predicted by the MGPT model.
easy and hard refer to different core configurations.  $\Delta$ is the
hard--easy core energy difference.}
\label{table:MGPT5x39x5}
\end{table*}

\begin{table*}
\begin{tabular}{c||c|c|c|c|c}
& \multicolumn{3}{c|}{Mo} & \multicolumn{2}{|c}{Ta}\\
\hline
$E_c$ (eV/$b$) & MGPT & AI$^*$ & AI & MGPT\cite{MGPTTa} & AI\\
\hline
hard & 2.57 & 2.94 & -- & -- & 0.91 \\
\hline
easy & 2.35 & 2.86 & 2.64 & -- & 0.86 \\
\hline
\hline
$\Delta$ & 0.22 & 0.08 & -- & 0.14 & 0.05
\end{tabular}
\caption{Core energies for $r_c=2b$ for fully relaxed {\em ab initio}
cores (AI) and interatomic (MGPT) cores in the 5$\times$3 cell.
AI$^*$ refers to {\em ab initio} core energies computed based on
relaxed MGPT configurations as the {\em ab initio} Mo hard core is
unstable.  Ref.\protect\cite{MGPTTa} only reported $\Delta$ for Ta.}
\label{table:compareabiMGPT}
\end{table*}


\begin{thebibliography}{0}
\bibitem{Schmid}{E. Schmid, Proc. Int. Congr. Appl. Mech., 342
(1942).}
\bibitem{Vitek}{V. Vitek, Cryst. Lattice Defects {\bf 5} (1974), and
references therein.}
\bibitem{Duesbery}{M.S. Duesbery, {\em Dislocations 1984}, (CNRS,
Paris 1984), p. 131.}
\bibitem{Hirsch}{Hirsch, P. B., Proc. 5th Int. Conf. Crystallography,
Cambridge University, 139 (1960).}
\bibitem{MGPT}{W. Xu and J. A. Moriarty, Phys. Rev. B 54, 6941 (1996).}
\bibitem{MGPTTa}{J. A. Moriarty et al., J. Engr. Mater. and Tech. {\bf 121} 120 (1999).}
\bibitem{Duesbery2}{Z.S. Basinski and M.S. Duesbery, in {\em
Dislocation Modeling of Physical Systems}, eds. M.F. Ashby {\em et
al.}, (Pergamon, Oxford 1980), p. 273.}
\bibitem{RMP}{M.C. Payne, M.P. Teter, D.C. Allen, T.A. Arias, and
J.D. Joannopoulos, Rev. Mod. Phys. {\bf 64}, 1045 (1992).}
\bibitem{PerdewZunger}{J. Perdew and A. Zunger, Phys. Rev. B. {\bf
23}, 5048 (1981).}
\bibitem{CeperlyAlder}{D. M. Ceperly and B. J. Alder, Phys. Rev. Lett
{\bf 45}, 566 (1980).}
\bibitem{KB}{L. Kleinmann and D. M. Bylander, Phys. Rev. Lett. {\bf
48}, 1425 (1982).}
\bibitem{Rappe}{A. Rappe, K. Rabe, E. Kaxiras, and J. D. Joannopoulos,
Phys. Rev. B {\bf 41}, 2127 (1990).}
\bibitem{Tapot}{C. Woodward, S. Kajihara, and L.H. Yang, Phys. Rev. B
{\bf 57}, 13459 (1998).}
\bibitem{ACPRL}{T. A. Arias, M. C. Payne, and J. D. Joannopoulos,
Phys. Rev. Lett. {\bf 69}, 1077 (1992).}
\bibitem{DFT++}{S. Ismail-Beigi and T. A. Arias, Journal of
Computational Chemistry, to be published.}
\bibitem{PayneSi}{J. R. K. Bigger et al, Phys. Rev. Lett. {\bf 69}
2224 (1992).}
\bibitem{HirthLothe}{J. P. Hirth and J. Lothe, {\em Theory of
Dislocations}, John Wiley \& Sons, Inc.: New York, 1982.}
\bibitem{Head}{A. K. Head, Phys. Stat. Solidi {\bf 5}, 51 (1964);
Phys. Stat Solidi {\bf 6}, 461 (1964).}
\bibitem{MGPT2}{W. Xu and J. A. Moriarty, Comput. Mater. Sci. 9, 348
(1998).}
\bibitem{Sigle}{W. Sigle, Phil. Mag. A {\bf 79}, 1009 (1999).}
\end{thebibliography}
\end{document}